\documentclass[12pt]{article} 
\usepackage[utf8]{inputenc} 
\usepackage[T1]{fontenc} 
\usepackage{amsmath,amssymb,bbm,bm,commath,mathtools,slashed,wasysym,xfrac} 
\usepackage{xcolor} 
\usepackage{graphicx} 
\graphicspath{{figures/}}
\usepackage{xparse} 
\usepackage{cite} 

\usepackage{silence}

\usepackage{sectsty} 
\WarningsOn[latex] 
\usepackage{geometry} 
\usepackage{fancyhdr} 
\usepackage{setspace} 
\usepackage[nottoc,notlof,notlot]{tocbibind} 
\usepackage[titles]{tocloft} 
\usepackage[pagebackref=true]{hyperref} 
\hypersetup{bookmarksnumbered=true, bookmarksopen=true, bookmarksopenlevel=2, breaklinks=true, citecolor=blue, colorlinks=true, hypertexnames=true, linkcolor=red, linktoc=page, pdfborder={0 0 0}, pdfstartview=FitH, pdfauthor={DJain}, pdfsubject={Quantization of Deformed QM}, plainpages=false, unicode=true, urlcolor=cyan}

\renewcommand*{\backrefalt}[4]{%
  \ifcase #1 \relax 
  \or {\footnotesize \raisebox{0.15em}{$\uparrow^{#2}$}} 
  \else {\footnotesize \raisebox{0.15em}{$\uparrow^{#2}$}} 
  \fi%
}

\DeclareUnicodeCharacter{0393}{\Gamma}
\DeclareUnicodeCharacter{0394}{\Delta}
\DeclareUnicodeCharacter{0398}{\Theta}
\DeclareUnicodeCharacter{039B}{\Lambda}
\DeclareUnicodeCharacter{039E}{\Xi}
\DeclareUnicodeCharacter{03A0}{\Pi}
\DeclareUnicodeCharacter{03A3}{\Sigma}
\DeclareUnicodeCharacter{03A5}{\Upsilon}
\DeclareUnicodeCharacter{03A6}{\Phi}
\DeclareUnicodeCharacter{03A8}{\Psi}
\DeclareUnicodeCharacter{03A9}{\Omega}
\DeclareUnicodeCharacter{03B1}{\alpha}
\DeclareUnicodeCharacter{03B2}{\beta}
\DeclareUnicodeCharacter{03B3}{\gamma}
\DeclareUnicodeCharacter{03B4}{\delta}
\DeclareUnicodeCharacter{03B5}{\epsilon}
\DeclareUnicodeCharacter{03B6}{\zeta}
\DeclareUnicodeCharacter{03B7}{\eta}
\DeclareUnicodeCharacter{03B8}{\theta}
\DeclareUnicodeCharacter{03D1}{\vartheta}
\DeclareUnicodeCharacter{03B9}{\iota}
\DeclareUnicodeCharacter{03BA}{\kappa}
\DeclareUnicodeCharacter{03BB}{\lambda}
\DeclareUnicodeCharacter{03BC}{\mu}
\DeclareUnicodeCharacter{03BD}{\nu}
\DeclareUnicodeCharacter{03BE}{\xi}
\DeclareUnicodeCharacter{03C0}{\pi}
\DeclareUnicodeCharacter{03C1}{\rho}
\DeclareUnicodeCharacter{03C3}{\sigma} 
\DeclareUnicodeCharacter{03C4}{\tau}
\DeclareUnicodeCharacter{03C5}{\upsilon}
\DeclareUnicodeCharacter{03C6}{\phi}
\DeclareUnicodeCharacter{03D5}{\varphi}
\DeclareUnicodeCharacter{03C7}{\chi}
\DeclareUnicodeCharacter{03C8}{\psi}
\DeclareUnicodeCharacter{03C9}{\omega}
\DeclareUnicodeCharacter{00B0}{^{\circ}}
\DeclareUnicodeCharacter{00B1}{\pm}
\DeclareUnicodeCharacter{00B7}{\cdot}
\DeclareUnicodeCharacter{00BD}{\tfrac{1}{2}}
\DeclareUnicodeCharacter{00D7}{\times}
\DeclareUnicodeCharacter{2020}{\dagger}
\DeclareUnicodeCharacter{210F}{\hbar}
\DeclareUnicodeCharacter{2190}{\leftarrow}
\DeclareUnicodeCharacter{2192}{\rightarrow}
\DeclareUnicodeCharacter{2194}{\leftrightarrow}
\DeclareUnicodeCharacter{21D0}{\Leftarrow}
\DeclareUnicodeCharacter{21D2}{\Rightarrow}
\DeclareUnicodeCharacter{21D4}{\Leftrightarrow}
\DeclareUnicodeCharacter{2202}{\partial}
\DeclareUnicodeCharacter{2207}{\nabla}
\DeclareUnicodeCharacter{2208}{\in}
\DeclareUnicodeCharacter{220F}{\prod}
\DeclareUnicodeCharacter{222B}{\int}
\DeclareUnicodeCharacter{222E}{\oint}
\DeclareUnicodeCharacter{2211}{\sum}
\DeclareUnicodeCharacter{2213}{\mp}
\DeclareUnicodeCharacter{221E}{\infty}
\DeclareUnicodeCharacter{2243}{\simeq}
\DeclareUnicodeCharacter{2248}{\approx}
\DeclareUnicodeCharacter{2260}{\neq}
\DeclareUnicodeCharacter{2261}{\equiv}
\DeclareUnicodeCharacter{2264}{\leq}
\DeclareUnicodeCharacter{2265}{\geq}
\DeclareUnicodeCharacter{226A}{\ll}
\DeclareUnicodeCharacter{226B}{\gg}
\DeclareUnicodeCharacter{2282}{\subset}
\DeclareUnicodeCharacter{2295}{\oplus}
\DeclareUnicodeCharacter{2297}{\otimes}
\DeclareUnicodeCharacter{22EF}{\cdots}
\DeclareUnicodeCharacter{25A1}{\square}
\DeclareUnicodeCharacter{266A}{\eighthnote}
\DeclareUnicodeCharacter{266B}{\twonotes}
\DeclareUnicodeCharacter{0212B}{\AA}

\definecolor{title}{rgb}{0.5,0.3,0.7}
\definecolor{abst}{rgb}{0.366,0.366,0.266}
\definecolor{sect}{rgb}{0.5,0.1,0.5}
\definecolor{ssect}{rgb}{0.5,0.05,0.25}
\definecolor{sssect}{rgb}{0.5,0.025,0.125}
\definecolor{appsect}{rgb}{0.5,0.1,0.5}
\definecolor{ref}{rgb}{0.5,0.3,0.5}
\definecolor{orcidlogocol}{HTML}{A6CE39}
\newcommand{\Title}[1] {\title{\color{title}\Huge #1}}
\newcommand{\TPheader}[3] {\date{}\maketitle\thispagestyle{fancy}\pagenumbering{alph}\lhead{#1}\chead{#2}\rhead{#3}\cfoot{}}

\newcommand{\Abstract}[1] {\begin{abstract}\normalsize #1 \end{abstract}}

\sectionfont{\color{sect}}
\subsectionfont{\color{ssect}}
\subsubsectionfont{\color{sssect}}
\renewcommand{\appendix}{\setcounter{section}{0}\sectionfont{\color{appsect}}\renewcommand{\thesection}{\Alph{section}}\renewcommand*{\theHsection}{app.\the\value{section}}} 
\newcommand\references[1]{\sectionfont{\color{ref}}\bibliographystyle{hephys}\bibliography{#1}}

\newcommand\eqs[1] {\begin{align}#1\end{align}}

\newcommand\eqst[1] {\begin{multline}#1\end{multline}}

\newcommand\eqsa[1] {\equ{\begin{aligned}#1\end{aligned}}}
\newcommand\eqsg[1] {\equ{\begin{gathered}#1\end{gathered}}}
\newcommand\equ[1] {\begin{equation}#1\end{equation}}

\newcommand\snmat[1] {\begin{smallmatrix}#1\end{smallmatrix}}
\newcommand\spmat[1] {\(\begin{smallmatrix}#1\end{smallmatrix}\)}

\newcommand\half {\tfrac{1}{2}}

\renewcommand\( {\left(}
\renewcommand\) {\right)}

\newcommand\N {{\mathcal N}} 
\renewcommand\O {{\mathcal O}}

\newcommand\nn {\nonumber\\}

\numberwithin{equation}{section} 
\begin{document}
\Title{Analyzing Uniform WKB for Deformed QM\\[-3mm] {\Large Or}\\[-2mm] How Not to Quantize the SW Curve}

\author{\href{mailto:dharmesh.jain@outlook.com}{Dharmesh Jain}\,\href{https://orcid.org/0000-0002-9310-7012}{\includegraphics[scale=0.0775]{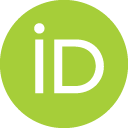}}
\smallskip\\
\emph{Rajasthan, India}
}

\TPheader{}{April 01, 2026}{} 

\Abstract{We uncover an inconsistency in the uniform WKB quantization of deformed quantum mechanics.
}

\tableofcontents 
\pagenumbering{arabic}

\section{Introduction}
The quantization of the Seiberg-Witten (SW) curve for $\N=2$ $SU(N)$ SYM theory\cite{Seiberg:1994rs} leads to a Hamiltonian of the following form\cite{Lawrence:1997jr,Codesido:2015dia}
\equ{H=Λ\(\cosh\(\frac{p}{\sqrt{mΛ}}\)-1\) +V(x)\,.
\label{defDQM}}
Here $[x, p] = i\hbar$ and $V(x)$ is a polynomial of degree $N$ in $x$. The spectral problem for this Hamiltonian has been analyzed from the perspective of Topological String/Spectral Theory (TS/ST) correspondence in \cite{Grassi:2018bci,Francois:2025nmq}. As a result, exact quantization conditions and wavefunctions for this Hamiltonian have been conjectured/obtained. 

The $Λ→∞$ limit of the Hamiltonian in \eqref{defDQM} gives the usual Schr\"odinger Hamiltonian for a particle in the potential $V(x)$
\equ{H_{S}=\frac{p^2}{2m} +V(x)\,.
\label{defQM}}
Thus, the Hamiltonian of type \eqref{defDQM} with its ``$\cosh p\,-1$'' kinetic term, a deformation from the ordinary quantum mechanical Hamiltonian \eqref{defQM}, is termed deformed quantum mechanical Hamiltonian. Inspired by the work of \cite{Grassi:2018bci}, it is natural to ask if the conventional exact JWKB approach for ordinary quantum mechanics (QM) \cite{Silverstone:1985hjs} can be suitably generalized to directly obtain analogous results for the case of deformed QM. An effort to answer this question was undertaken in the thesis \cite{Baldino2023}. It turns out to be a promising undertaking but it rests on a crucial conjecture and unfortunately, we believe that that conjecture is not true.

In what follows, we review the key initial setup of uniform WKB approach for ordinary QM following \cite{Marino:2021lne}. Then we try to follow this setup as closely as possible for the deformed QM and prove that we come up against an inconsistency, rendering the promising approach not so.

\section{Uniform WKB for Ordinary QM}
The WKB method is an approximation technique to solve the eigenvalue problem $H_Sψ=Eψ$, or equivalently
\equ{-\frac{ℏ^2}{2m}ψ''(x) +V(x)ψ(x)=Eψ(x)\,.
\label{origSE}}
The uniform WKB method begins with the following ansatz for the wavefunction
\equ{ψ(x) =\frac{1}{\sqrt{φ'(x)}}f\(φ(x)\).
\label{psiansatz}}
The usefulness of this ansatz lies in the fact that $f'(φ)$ terms cancel out and the function $f\(φ\)$ satisfies the following second-order differential equation (note its similarity to the original Schrödinger equation \eqref{origSE})
\equ{ℏ^2f''(φ) +Π(φ)^2f(φ)=0\,.
\label{eqfphi}}
The function $Π(φ)$ in terms of $φ(x)$ is defined by the following nonlinear differential equation
\equ{Π(φ)^2φ'(x)^2 +\frac{ℏ^2}{2}\{φ(x),x\} =2m\(E-V(x)\),
\label{eqphix}}
where $\{φ(x),x\}$ is the Schwarzian derivative (whose explicit form is not relevant for our purposes, see \cite{Marino:2021lne} for details).

The equation \eqref{eqfphi} for $f(φ)$ can be ``exactly'' solved by choosing an appropriate $Π(φ)^2$. For example, $Π(φ)^2=1$ reproduces the standard WKB method with $f_±(φ)=e^{±\frac{i}{ℏ}φ}$. This choice has the drawback of breaking down near the turning points of the potential, so $Π(φ)^2=φ$ is a better choice when the potential behaves linearly near the turning points, leading to $f(φ)=a\,\text{Ai}\big(-ℏ^{-\frac{2}{3}}φ\big) +b\,\text{Bi}\big(-ℏ^{-\frac{2}{3}}φ\big)$. Note that $f(φ)$ is analogous to the wavefunction $ψ(x)$ for linear potential ($V(x)=x$). This ``connection'' to the linear problem is heavily exploited to get the solution for generic potential in the uniform WKB approach, but we will not go into those details here. The linear choice $Π(φ)^2=φ$, in turn, gives a nontrivial equation \eqref{eqphix} for $φ(x)$ that can, in principle, be solved order-by-order in $ℏ$. Thus, another way to think about \eqref{eqphix} is as a consistency condition for the existence of $φ(x)$ ensuring the validity of the original ansatz \eqref{psiansatz}. An analogous condition was conjectured to hold true for the uniform WKB method when applied to the deformed QM case \cite{Baldino2023}. We analyse it now.

\section{Uniform WKB for Deformed QM}
Let us now return to \eqref{defDQM} and set up the spectral problem for deformed QM ($Hψ=Eψ$):
\equ{½\(\psi(x+i\hbar) + \psi(x-i\hbar)\) -ψ(x) +V(x)ψ(x) = Eψ(x)\,,
\label{diffeqDQM}}
where we set $\frac{ℏ}{\sqrt{mΛ}}→ℏ$ for convenience. Note that this is a second-order difference equation and not a differential one, so let us become familiar with some difference calculus now\cite{Ghoshal:2015sg}.

We define a difference operator $D_{ℏ}$ and a shift operator $T_ℏ$ as follows
\eqs{D_ℏψ(x) &= \tfrac{1}{iℏ}(ψ(x+iℏ) -ψ(x)) \\
T_ℏψ(x) &≡e^{iℏ∂_x}ψ(x) =ψ(x+iℏ)\,.
}
At the operator level, $D_ℏ =\frac{1}{iℏ}(T_ℏ-I)$ and $D_{-ℏ}=\tfrac{1}{iℏ}(I-T_{-ℏ})$ (with $I$ being the identity operator) such that their products satisfy the following interesting relations
\equ{T_ℏD_{-ℏ} =D_{ℏ}\,;\quad D_ℏT_{-ℏ}=D_{-ℏ}\,;\quad D_ℏD_{-ℏ} =-\tfrac{1}{ℏ^2}(T_ℏ +T_{-ℏ} -2I)\,.
\label{Dh2op}}
The product and chain rules for the difference operator will prove useful and are as follows
\eqs{D_ℏ[f(x)g(x)] &=\(D_ℏf(x)\)g(x) +\(T_ℏf(x)\)\(D_ℏg(x)\) \\
D_ℏf(φ(x)) &=D_ℏφ(x)D_{ℏD_ℏφ(x)}f(φ)\,,
}
where the convoluted subscript of the rightmost $D$ acts on $φ$ and not $x$.

Finally, we conclude from \eqref{Dh2op} that $\frac{iℏ}{2}(D_ℏ -D_{-ℏ})≡-\frac{ℏ^2}{2}D_ℏD_{-ℏ} ≡\cosh p-1$ is the ``kinetic operator'' of the deformed QM, so in analogy with ordinary QM, we can rewrite \eqref{diffeqDQM} as
\equ{-\tfrac{ℏ^2}{2}D_ℏD_{-ℏ}ψ(x) +V(x) ψ(x)=Eψ(x)\,.
\label{DQMgenpot}}

\subsection{The Linear Problem}
The deformed quantum mechanical problem with a linear potential $V(x)=gx$ (with $g>0$) is given by the following rewriting of the difference equation in \eqref{diffeqDQM}
\equ{\psi(x+i\hbar) + \psi(x-i\hbar) = 2\(1+E-gx\)\psi(x)\,,
\label{DQMlinpot}}
and it is well known \cite{Watson1922} that this is solved by the Bessel functions
\equ{ψ(x,ℏ) = A(q,ℏ)J_{\frac{y}{a}}\(\tfrac{1}{a}\) +B(q,ℏ)Y_{\frac{y}{a}}\(\tfrac{1}{a}\).
\label{DQMlinpotsol}}
Here $y=1+E-gx$, $a=iℏ$ and $q=e^{-\frac{2πiy}{a}}$, such that $A$ and $B$ coefficients can be expanded (in general) as $A(q,ℏ)=∑_{n=-∞}^∞a_n(ℏ)q^n$ and $B(q,ℏ)=∑_{n=-∞}^∞b_n(ℏ)q^n$.

\subsection{The General Problem}
For the purpose of implementing the uniform WKB algorithm, we consider an ansatz for the wavefunction as follows
\equ{ψ(x)=e^{f_o(φ(x))}f(φ(x))\,,
}
where $f_o(φ)$ should involve $D_{±ℏ}φ(x)$ in some non-trivial way (like \cite[(3.2.30)]{Baldino2023}). Acting with $D_ℏD_{-ℏ}$ on this ansatz, we generate
\begingroup
\allowdisplaybreaks
\eqs{D_ℏD_{-ℏ}\big(e^{f_o(φ)}f(φ(x))\big) &=\big(D_ℏD_{-ℏ}e^{f_o(φ)}\big)f(φ(x)) +e^{f_o(φ)}\(D_ℏD_{-ℏ}f(φ(x))\) \nn
&\quad +\big(D_ℏe^{f_o(φ)}\big)D_ℏf(φ(x)) +\big(D_{-ℏ}e^{f_o(φ)}\big)D_{-ℏ}f(φ(x)) \nn
&=\big(D_ℏD_{-ℏ}e^{f_o(φ)}\big)f(φ(x)) \nn
&\quad +\tfrac{1}{iℏ}\big(T_ℏe^{f_o(φ)}\big)D_{ℏ}φ(x)D_{ℏD_{ℏ}φ(x)}f(φ) \nn
&\quad -\tfrac{1}{iℏ}\big(T_{-ℏ}e^{f_o(φ)}\big)D_{-ℏ}φ(x)D_{-ℏD_{-ℏ}φ(x)}f(φ)\,.
\label{actdiffeq}}
\endgroup
Substituting this in \eqref{DQMgenpot}, we get
\eqst{\tfrac{iℏ}{2}\big(T_ℏe^{f_o(φ)}\big)D_{ℏ}φ(x)D_{ℏD_{ℏ}φ(x)}f(φ) -\tfrac{iℏ}{2}\big(T_{-ℏ}e^{f_o(φ)}\big)D_{-ℏ}φ(x)D_{-ℏD_{-ℏ}φ(x)}f(φ) \\
=\tfrac{ℏ^2}{2}\big(D_ℏD_{-ℏ}e^{f_o(φ)}\big)f(φ(x)) +\(E-V(x)\)e^{f_o(φ(x))}f(φ(x)).
}
For the uniform WKB approach to work here (as in the ordinary QM case), we expect the function $f(φ)$ to satisfy the original difference equation with a linear potential (in $φ$). So we demand that the LHS above can be written as
\equ{\half e^{\tilde{f}(φ)}\(f(φ+iδφ)+f(φ-iδφ) -2f(φ)\),
}
such that we get the following equation for $f(φ)$:
\eqsg{f(φ+iδφ)+f(φ-iδφ) =2\tilde{Π}(φ)f(φ)\,, \\
\tilde{Π}(φ) =1+e^{-\tilde{f}(φ)}\left\{\tfrac{ℏ^2}{2}D_ℏD_{-ℏ}+E-V(x)\right\}e^{f_o(φ)}\,,
\label{diffequniWKB}}
where setting $\tilde{Π}(φ)=1+E-φ$ leads us to an equation similar to \eqref{DQMlinpot} for the linear potential. After this, we expect the uniform WKB algorithm to take over and lead us to the connection formulae and the exact quantization conditions as claimed in \cite{Baldino2023}. But before that, we need to make sure that the demand we made above can be fulfilled, namely,
\eqst{iℏ\big(T_ℏe^{f_o(φ)}\big)D_{ℏ}φ(x)D_{ℏD_{ℏ}φ(x)}f(φ) -iℏ\big(T_{-ℏ}e^{f_o(φ)}\big)D_{-ℏ}φ(x)D_{-ℏD_{-ℏ}φ(x)}f(φ) \\
\stackrel{?}{=}e^{\tilde{f}(φ)}\(f(φ+iδφ)+f(φ-iδφ) -2f(φ)\).
\label{conscond}}

\paragraph{Claim:} The consistency condition \eqref{conscond} for uniform WKB in the deformed QM case can not be satisfied.\\[2mm]
{\it Proof:} Since we would like $Λ→∞$ (or $ℏ→0$ in our convenient notation) limit to exist, we can expect the following series expansions
\eqsa{e^{f_o(φ)} &=∑_{n=0}^∞ g_n\big(φ'(x),⋯,φ^{(n+1)}(x)\big)ℏ^n \\
e^{\tilde{f}(φ)} &=∑_{n=0}^∞ \tilde{g}_n\big(φ'(x),⋯,φ^{(n+1)}(x)\big)ℏ^n
}
as well as $δφ=∑_{n=1}^∞ δφ_n ℏ^n$.\footnote{We can not have $δφ_n$'s being $φ^{(n)}(x)$-dependent, as that would preclude $f(φ)$ being the Bessel functions.} Plugging these in \eqref{conscond}, we get (ignoring $\O(ℏ^3)$)
\begingroup
\allowdisplaybreaks
\eqs{\O(ℏ^2):\; & f'(φ)\(g_0(φ')+2φ'g'_0(φ')\)φ'' +f''(φ) g_0(φ'){φ'}^2 = f''(φ)\tilde{g}_0(φ')δφ_1^2\nn
\O(ℏ^4):\; & f'(φ)\(⋯\) +f''(φ)\(⋯\) +f'''(φ)\(\tfrac{1}{2}g_0(φ')+\tfrac{1}{3}φ'g'_0(φ')\){φ'}^2φ'' +\tfrac{1}{12} f^{(4)}(φ)g_0(φ'){φ'}^4 \nn
& =f''(φ)\(⋯\) +\tfrac{1}{12}f^{(4)}(φ)\tilde{g}_0(φ')δφ_1^4\,.
}
\endgroup
At $\O(ℏ^2)$, we need to set $\tilde{g}_0(φ')δφ_1^2=g_0(φ')φ'^ 2$ and $g_0(φ')+2φ'g'_0(φ')=0$. The later equation is solved by $g_0(φ')=\frac{c_0}{\sqrt{φ'}}$ (consistent with the ordinary QM case). However, this leads to a problem at $\O(ℏ^4)$ for the nonzero coefficient of $f'''(φ)$, which must be set to 0 leading to $g_0(φ')=0$, but this is inconsistent!  \hfill Q.E.D.$\blacksquare$

This means we cannot use the asymptotics of Bessel functions $f(φ)$ to map out how $e^{±\frac{S}{iℏ}}$ behave in different regions via uniform WKB ansatz as done in section \cite[3.2.2]{Baldino2023}.

\section{Further Issues}
Even if the above claim turns out to be ``unfounded'', the results of the thesis \cite{Baldino2023} still need some issues ironed out in their derivations and/or applications. We enumerate such issues in no particular order (the equation numbers below refer to equations in the thesis):
\begin{enumerate}
\item Start with (3.2.26), $-\frac{1}{ℏV'}∫_{φ_0}^1\cos^{-1}tdt =\frac{1}{ℏ}∫_x^{x_0}\cos^{-1}(E-V(t)+1)dt=\frac{T}{ℏ}$, and shift $T$ by $-2πn(x-x_0)$ leading to a factor of $q_{x_0}^n$. This amounts to a shift by $+2πn$ in the integrand of LHS (due to the limits) leading to $(q^{(φ)})^n$, without the sign flip, unlike in (3.2.29). 
\item Row multiplication of coefficient vector and connection matrix $\spmat{A(q) & B(q)}M(q)_{2×2}$ is used in equations like (3.2.32) of section 3.2.3, whereas column multiplication $M(q)_{2×2}$ $\big(\snmat{A(q) \\ B(q)}\big)$ is used in chapter 4 as in (4.1.12). We believe this is inconsistent with how the transition matrices were constructed in the first place.
\item Having done the Stokes analysis for $y<-1$ following \cite{Watson1922} and then repeating steps similar to those in section 3.2.2, we find that the conjugation equation (3.2.59), which relates the transition matrices for type $-1$ turning point to those for type $1$ turning points, does not hold! {\it These details are beyond the scope of this note, but see the included \href{HowNot2TMissue.nb}{Mathematica notebook} for the explicit matrices.}
\end{enumerate}

\section*{\centering Acknowledgements}
DJ thanks Sujay Ashok for heavy exposure to the works of Airy, Bessel, Borel, Stokes, et al.

\references{specprob.bib}

\end{document}